\documentstyle[mnras_cite]{mn}
\def\plotfiddle#1#2#3#4#5#6#7{\centering \leavevmode
\vbox to#2{\rule{0pt}{#2}}
\includegraphics{#1}}
\begin{document}
\title[Mass estimation in the outer regions of galaxy clusters]
{Mass estimation in the outer regions of galaxy clusters}
\author[A. Diaferio]{Antonaldo Diaferio\\
Max-Planck-Institut f\"ur Astrophysik, Karl-Schwarzschild-Str. 1, D-85740, Garching bei M\"unchen, Germany}

\maketitle
\begin{abstract}

We present a technique for estimating the mass in the outskirts of galaxy clusters where
the usual assumption of dynamical equilibrium is not valid. The method
assumes that clusters form through hierarchical clustering and
requires only galaxy redshifts and positions on the sky. 
We apply the method to dissipationless cosmological $N$-body simulations where
galaxies form and evolve according to semi-analytic modelling.
The method recovers the actual cluster mass profile within a factor of two to
several megaparsecs from the cluster centre. This error originates from 
projection effects, sparse sampling, and contamination by foreground and background galaxies. 
In the absence of velocity biases, this method can provide an estimate of the mass-to-light
ratio on scales $\sim 1-10h^{-1}$ Mpc where this quantity is still poorly known. 

\end{abstract}


\section{INTRODUCTION}

At scales $\la 1h^{-1}$ Mpc\footnote{We use the Hubble constant
$H_0=100h$ km s$^{-1}$ Mpc$^{-1}$ throughout.}, 
dynamics of galaxy groups and of the central region of clusters give
mass-to-light ratios $M/L\sim 20-30\%$ of that required to close the Universe (e.g.
\pcite{carlberg96}; \pcite{ramella97}). Observations of light curves of
Type Ia supernovae at redshifts $z< 1$ confirm that there might not be indeed enough
mass to close the Universe (\pcite{riess98}; \pcite{garnavich98};
\pcite{perlmutter98}). 
On the other hand, from the analysis of velocity fields on mildly non-linear scales 
($\ga 10h^{-1}$ Mpc),
we can measure the quantity $\beta=\Omega_0^{0.6}/b$ (e.g. \pcite{strauss95}) 
where $\Omega_0$ is the mean mass density of the Universe and $b$, assumed independent of scale,
is the bias parameter, the ratio between the galaxy and the mass overdensity fields.
Recent estimates yield $\beta\sim 0.5-1.1$ (e.g. \pcite{dekel97}; \pcite{schmoldt99}). 
If galaxies cluster more than mass ($b>1$), this result implies large values
of $\Omega_0$.

At intermediate scales, $\sim 1-10h^{-1}$ Mpc, 
estimating the mass-to-light ratio is particularly difficult.
These scales correspond to the immediate surroundings of rich clusters,
the largest non-linear systems forming at the present epoch. Matter is falling onto
the cluster for the first time. Therefore, 
neither dynamical equilibrium nor linear theory are valid
descriptions of the dynamics of this falling matter: at these scales, clusters are
already in the non-linear regime although not yet virialized.

We can derive constraints on $\Omega_0$ from the dynamics of these 
infall regions on a statistical basis,
by modelling the galaxy-cluster cross correlation function (e.g. \pcite{croft99}).
On a single cluster basis, however, using the spherical infall model
to extract information
on $\Omega_0$ (\pcite{regos89}) seems impractical, because in hierarchical clustering scenarios
random motions are an important component of the velocity field 
(van Haarlem \& van de Weygaert 1993). 
Nonetheless, the galaxy distribution in redshift
space still contains information
about the cluster mass (\pcite{diaferio97}, DG hereafter). 

DG suggest a method for measuring the mass profile at radii {\it larger} than the virial radius
of galaxy clusters. DG show that, 
if the redshift space coordinates of the dark matter particles were measurable,
we could estimate the mass profile with a $\sim 30\%$
accuracy, independently of the mass and the dynamical state of the halo.

Here, we present an operational method for extracting the mass profile
from the redshift space distribution of galaxies within real clusters.
This mass estimator relies on
kinematic data alone. Thus, the mass estimate is independent of the relative distribution
of mass and light, unless galaxies are not good tracers of the velocity field in the
infall region. Detailed modelling suggests that velocity bias is very weak on these scales
(\pcite{kauffmann98}; \pcite{Diaferio99}).
We apply the method to catalogues of galaxies formed and evolved using semi-analytic procedures
within the dark matter halos of dissipationless $N$-body simulations. 
We show that it  is indeed possible to measure the mass of clusters
to large radii. Because a few hundred galaxy redshifts are required
to give reliable results, this measurement is feasible only on rather massive clusters. 
This method enables us to estimate the value of 
the mass-to-light ratio on scales $\sim 1-10h^{-1}$ Mpc. 

Sections \ref{sec:rd} and \ref{sec:mass} review the 
DG interpretation of the galaxy density distribution
in the redshift diagram of clusters and the assumptions of their mass estimation
method.  Section \ref{sec:A} outlines the operational method of extracting the mass profile
from the redshift diagram, and 
Section \ref{sec:result} shows some applications to galaxy clusters extracted
from two cosmological models (Section \ref{sec:nbody}).
In Appendix \ref{app:centre} we also describe a method of locating the centre of a galaxy cluster
from a list of galaxies with redshift space coordinates.

\section{INTERPRETING THE REDSHIFT DIAGRAM}\label{sec:rd}

Observations provide three out of the six phase space coordinates of a galaxy.  Define
the redshift diagram as the plane $(r,v)$, where $r$ is the galaxy angular separation 
from the cluster centre and $v$ is its line-of-sight velocity relative to the cluster centre of mass. 

The spherical infall model predicts the existence of two curves in this plane 
where the galaxy number density is infinite (\pcite{regos89}). 
These caustics define a characteristic ``trumpet'' shape and 
enclose galaxies which are, in real space, at distances both smaller and larger 
than the turnaround radius from the cluster centre. Galaxies outside the
caustics are only at distances larger than the turnaround radius.
We define the amplitude ${\cal A}(r)$ of these caustics in redshift space as half
the difference between the upper and the lower caustic. The sphericall infall
model predicts a dependence of ${\cal A}(r)$ on the cosmological density
parameter $\Omega_0$ (\pcite{regos89}; DG). 

A clean contrast between the regions interior and exterior to the
caustics seem indeed evident in the redshift diagrams of
some real clusters (\pcite{quintana96}; \pcite{denhartog96};
Geller, Diaferio \& Kurtz 1999a; \pcite{geller98}).
However, it is a challenge to identify the location of the caustics 
reliably (e.g. \pcite{regos89}; \pcite{vanhaarlem93a}).
\scite{vanhaarlem93b} 
were the first to use 
$N$-body simulations of dark matter halos formed in hierarchical clustering scenarios to 
show  that (1) even in redshift diagrams of  
dark matter particles the caustics can be poorly determined, and (2)
the amplitude of the plausible caustics is usually larger than the amplitude 
predicted by the spherical infall model for a given cosmology.
They concluded correctly that the identification of caustics in cluster redshift diagrams is 
not a reliable method of estimating $\Omega_0$.

\scite{vanhaarlem93b} did not clarify, however, 
what ultimately produces the caustics in the redshift diagram of real clusters.
DG suggest a heuristic argument to explain both the presence
and the amplitude of the caustics. In hierarchical clustering
scenarios non-radial motions and substructure
obscure the pure radial motion predicted by the spherical infall model and
inflate the predicted amplitude of the caustics.  
How much do random motions inflate this amplitude? 
In the infall region of a cluster, a galaxy 
with a large enough velocity can escape the cluster gravitational
field in a time shorter than the age of the Universe.\footnote{As a rough estimate
of the astrophysical quantities involved, consider 
the gravitational field generated by a point of mass $M$.
The escape velocity at a distance $r$ from $M$ is
\begin{equation}
v_{\rm e}(r) \simeq 927\left(M\over 10^{14} h^{-1}{\rm M_\odot}\right)^{1/2}
 \left(r\over h^{-1}{\rm Mpc}\right)^{-1/2}
 {\rm km\ s}^{-1}.
\label{eq:escnote}
\end{equation}
In such a field, the deceleration is
\begin{equation}
 \dot v(r)  \simeq -440 \left(M\over 10^{14} h^{-1}{\rm M_\odot}\right) 
 \left(r\over h^{-1}{\rm Mpc}\right)^{-2} h\ {\rm km\ s}^{-1} {\rm Gyr}^{-1}.
\end{equation}
Recall that a velocity of $1000$ km s$^{-1}$ corresponds to $\sim 1$ Mpc Gyr$^{-1}$.}
We may therefore expect only a few escaping galaxies lying outside the caustic region in the
redshift diagram; this fact rapidly decreases the galaxy number density
outside the caustics. Thus, the appearence of the ``caustics'' is 
related to the local escape velocity rather than to the cosmological
density parameter.  In other words,
the caustic amplitude is a measure of the cluster gravitational potential
at a particular distance from the cluster centre.

Of course, we do not have a correlation between caustic amplitude and
cluster gravitational potential when a  
galaxy system falling onto the central cluster has a mass comparable to that of
the cluster; in this case,
the amplitude of the caustics is related to the gravitational potential
of this falling system rather than the cluster itself.

In order to relate the caustic amplitude ${\cal A}(r)$ to the cluster gravitational potential
$\phi(r)$
we proceed as follows (DG).\footnote{We restrict our argument to spherical symmetry because we
expect the azimuthal average to enhance
the appearence of the caustics in the redshift diagram.}
Regardless of the stability of the cluster, at any particular shell of radius
$r$ the maximum allowed velocity is the escape velocity
$v_{\rm e}^2(r)=-2\phi(r)$.
The line-of-sight component of this escape velocity sets the location and the 
amplitude of the caustics in the redshift diagram.

\begin{figure}
\plotfiddle{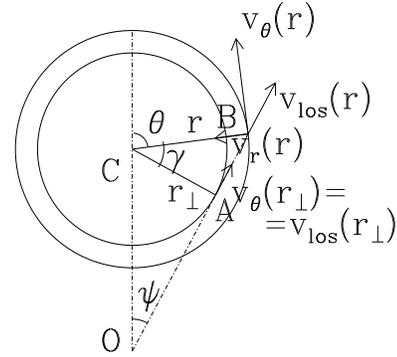}
           {0.2\vsize}              
           {90}                
           {29}                 
           {30}                 
           {100}               
           {-10}                
\caption{Geometry of the problem. An observer $O$ views the cluster centre $C$
along the line of sight $OC$ and the galaxies $A$ and $B$ along the line of sight
$OAB$ at angular separation $\psi$ from the cluster centre. The longitudinal angles
of $A$ and $B$ are $\theta+\gamma$ and $\theta$, respectively.}
\label{fig:scheme}
\end{figure}

We now compute this expected maximum observable velocity. 
The line-of-sight component of $ v_{\rm e}(r)$ 
at any given angular separation $r_\perp$ from the cluster centre is (see e.g. 
point $B$ in Fig. \ref{fig:scheme})
\begin{equation}
v_{\rm e,los}(r\vert r_\perp)= v_{{\rm e},\theta}(r) {r_\perp\over r} - v_{{\rm e},r}(r)
{(r^2-r_\perp^2)^{1/2}\over r},
\label{eq:scheme}
\end{equation}
where the longitudinal angle $\theta$ is the angle between the galaxy position
vector ${\bf r}$ centered on the cluster and the line-of-sight to the cluster centre.
The equation $\partial v_{\rm e,los}(r\vert r_\perp)/\partial r=0$ 
yields the maximum observable velocity along
the line-of-sight at fixed angular separation $r_\perp$.
The radius $r=r_M$, where this maximum occurs, generally differs from $r_\perp$ and
depends on the velocity profiles $v_{{\rm e},r}(r)$ and $v_{{\rm e},\theta}(r)$, i.e.
the gravitational potential $\phi(r)$ and the tidal field determining the non-radial
component of the velocity field.

We seek an expression for the expected amplitude ${\cal A}(r_\perp)$ 
of the caustics which is independent of the form of the escape
velocity profiles. 
Consider the cluster velocity anisotropy
parameter $\beta(r)=1-\langle v_\theta^2
+v_\phi^2\rangle/2\langle v_r^2\rangle$,
where $v_\phi$ is the azimuthal component of
a galaxy velocity ${\bf v}$, and the angle brackets indicate an average over the velocities of
all the galaxies within the volume $d^3{\bf r}$ centered on position ${\bf r}$. 
If the cluster rotation is negligible,
$\langle v_\theta^2+v_\phi^2\rangle=2\langle v_\theta^2\rangle=\langle v^2\rangle-
\langle v_r^2\rangle$ and
we can write $\langle v_{{\rm e}}^2(r_\perp)\rangle$, 
the mean of the square of the escape velocity on the shell of radius $r_\perp$,
by combining its line of sight component
$v_{\rm e,los}^2(r_\perp)=v_{{\rm e},\theta}^2(r_\perp)$
(equation \ref{eq:scheme}) and $\beta(r_\perp)$:
\begin{equation}
\langle v_{{\rm e},\theta}^2(r_\perp)\rangle=
\langle v_{\rm e}(r_\perp)^2\rangle {1-\beta(r_\perp)\over
3-2\beta(r_\perp)}.
\label{eq:esc}
\end{equation}
Now, define the quantities 
\begin{equation}
g(\beta)\equiv{3-2\beta(r_\perp)\over 1-\beta(r_\perp)}, \quad\quad
\phi_\beta(r_\perp) \equiv {2\vert\phi(r_\perp)\vert\over  g(\beta)};
\label {M0}
\end{equation}
$g(\beta)$ contains the information about the anisotropy of the velocity
field governed by the infall and the tidal field; $\phi_\beta(r_\perp)$ 
combines $g(\beta)$ with the potential $\phi(r_\perp)$ determined by the mass distribution alone.
By assuming $2\vert\phi(r_\perp)\vert\simeq\langle v_{\rm e}^2(r_\perp) \rangle$, we
can rewrite equation
(\ref{eq:esc}) as 
\begin{equation}
\phi_\beta(r_\perp) \simeq \langle v_{{\rm e},\theta}^2(r_\perp)\rangle
=\langle v_{{\rm e,los}}^2(r_\perp)\rangle.
\label{eq:phibeta}
\end{equation}

When $r_M\ne r_\perp$, 
$v_{\rm e,los}^2(r=r_M\vert r_\perp)> v_{\rm e,los}^2(r=r_\perp\vert r_\perp)$.
Thus, {\it a priori}, 
the {\it true} amplitude $\langle v_{\rm e,los}^2(r=r_M\vert r_\perp)\rangle$ 
of the caustics is larger than $\phi_\beta(r_\perp)$.
Moreover, the {\it measured} ${\cal A}(r_\perp)$ can be larger
or smaller than $\langle v_{\rm e,los}^2(r=r_M\vert r_\perp)\rangle$ depending
on the presence or absence of escaping galaxies in the redshift diagram.
However, comparisons with  clusters in $N$-body simulations indicate that
$\phi_\beta(r_\perp)$, computed with the full phase space information, agrees,
within the uncertainties, with the amplitude ${\cal A}^2(r_\perp)$ computed 
from the redshift diagram (DG; see also Section \ref{sec:result}).
We can therefore consider ${\cal A}^2(r_\perp)$ as a measure of $\phi_\beta(r_\perp)$
\begin{equation}
\phi_\beta(r_\perp)\rightarrow{\cal A}^2(r_\perp).
\end{equation}
The equation above represents our physical interpretation of the caustic
amplitude ${\cal A}(r_\perp)$ in the redshift diagram of clusters.

Equation (\ref{M0}) clearly shows that non-radial motion is the fundamental ingredient of this
interpretation of the caustic amplitude ${\cal A}(r_\perp)$. In fact, 
if orbits were radial, $\beta(r_\perp)=1$, and  equation (\ref{M0}) would predict 
an uncorrect $\phi_\beta(r_\perp)=0$. On the other hand, random motions dominate 
the velocity field of the virialized region of the cluster. 
Thus, this interpretation holds both in the 
central and in the infall region of the cluster.

In conclusion, in redshift diagrams of real systems, we expect a concentration
of galaxies around the cluster redshift at any fixed angular separation. Because of infall,
we expect a somewhat clear separation between falling galaxies and unrelated 
galaxies which are more distant than the turnaround radius.
Because of random motions, this separation is not as sharp as predicted by 
the spherical infall model, but it is still apparent, and it is indeed observed (\pcite{geller99a}).
Finally, the cluster gravitational potential determines the amplitude of the caustics.

\section{ESTIMATING THE INTERIOR MASS}\label{sec:mass}

Suppose we have a method to measure ${\cal A}(r)$ and $g(\beta)$.
By assuming ${\cal A}^2(r)=\phi_\beta(r)$, 
equation (\ref{M0}) readily yields, for the cumulative cluster 
mass $GM(<r)=r^2d\phi/dr$,
\begin{equation}
GM(<r) = -{r\over 2}{\cal A}^2(r)g(\beta)\left({d\ln{\cal A}^2\over d\ln r}
+{d\ln g\over d\ln r}\right).
\label {Mtrue}
\end{equation}
The two logarithmic derivatives in equation (\ref{Mtrue}) are comparable.
Recently, \acite{merrifield98} has shown that $\beta(r)$
could actually be measured by detecting $N\ga 50$ galaxy wakes in the X-ray emission of
clusters. Bent lobes of radio galaxies might also be used
to cross-check these measurements; these galaxies will not be useful for 
actual measurements, however, because the expected number of these galaxies  
is not larger than $\sim 5$ per cluster (\pcite{cress99}).
In any case, even if we had a measure of $\beta(r)$, a serious problem
occurs independently of the method
we use to estimate ${\cal A}^2(r)$: sparse sampling and background
and foreground galaxies will yield a very noisy ${\cal A}^2(r)$; therefore 
its differentiation is not practical even after substantial smoothing (DG).

Nevertheless, a measure of ${\cal A}(r)$ alone can still be
used to estimate the cluster mass.
DG suggest casting the relation between ${\cal A}(r)$ and $M(<r)$ in 
the form 
\begin{equation}
GM(<r) - GM(<r_0) =
\int_{r_0}^{r}  {\cal A}^2(x) {\cal F}_\beta (x)dx
\label {M1}
\end{equation}
where
\begin{eqnarray}
{\cal F}_\beta (r)&=&-2\pi G {\rho(r) r^2\over\phi(r)} {3-2\beta(r)\over 1-\beta(r)} \\
\label{M1.0}
 & \equiv & {\cal F}(r) g(\beta),
\label{M1.1}
\end{eqnarray}
$\rho(r)$ is the cluster mass density profile and
\begin{equation}
\phi(r) = -{GM(<r)\over r} - 4\pi G \int_r^\infty \rho(x) x dx
\label {M4}
\end{equation}
is the gravitational potential generated by the cluster.

Equation (\ref{M1}) integrates the estimated ${\cal A}^2(r)$
rather than differentiating it, thus averaging out its un-physical rapid variations.
However, the crucial property of equation (\ref{M1}) is that 
the function ${\cal F}_\beta(r)$ is slowly varying at large radii 
in hierarchical clustering scenarios, as we will show below. 
This property is of major importance, because we may assume
${\cal F}_\beta(r)={\rm const}\equiv {\cal F}_\beta$. Equation ({\ref{M1}) then becomes
\begin{equation}
GM(<r) - GM(<r_0) = {\cal F}_\beta \int_{r_0}^r  {\cal A}^2(x)dx.
\label {M2}
\end{equation}
Thus, {\it we make the mass estimation method independent of any scale length},
implicitly assumed in the function ${\cal F}_\beta(r)$, although
still dependent on the unknown ${\cal F}_\beta$. Of the three equations (\ref{Mtrue}),
(\ref{M1}) and (\ref{M2}), only equation (\ref{M2}) turns out to be of 
practical relevance. We emphasize that equation (\ref{M2}) is justified
only in hierarchical clustering scenarios. 

Note that the similarity between equation (\ref{M1}) and 
equation (\ref{M2}) might suggest 
using an iterative procedure: equation (\ref{M2})
provides the first mass profile from which we can
estimate ${\cal F}_\beta(r)$ and we can then use equation (\ref{M1}) 
until convergence is achieved. Unfortunately, ${\cal F}_\beta(r)$ contains information
on both the mass distribution and the velocity field (equation \ref{M1.0}),
whereas equations (\ref{M1}) and (\ref{M2}) return information on the
mass profile alone. Therefore, we should still make an assumption about $g(\beta)$.
However, even in this case, the improvement in  the mass estimate 
is likely to be insufficient to reduce the uncertainty 
which is introduced by projection effects. These effects actually dominate the problem,
as we will show in Section \ref{sec:result}.

To justify ${\cal F}_\beta(r)={\cal F}(r)g(\beta)={\rm const}$ in
hierarchical clustering cosmogonies, first
consider the function ${\cal F}(r)=-2\pi G \rho(r) r^2/ \phi(r)$,
and a density profile approximated piecewise
by power laws $\rho(r) \propto r^{-\alpha(r)}$. 
Provided the mass is finite at $r=0$ ($\alpha(r)< 3$, when $r\to 0$),
and $\phi(r)$ converges at large radii ($\alpha(r)>2$, when $r\to\infty$),
we obtain ${\cal F}(r)\propto r^{2-\alpha(r)}/[M(<r)r^{-1}+r^{2-\alpha(r)}]$.
At large radii, ${\cal F}(r)$ behaves differently depending 
on the behavior of $M(<r)\propto r^{3-\alpha(r)}$. 
If the mass is finite ($\alpha(r)>3$, when $r\to\infty$),
${\cal F}(r)$ decreases as a power law,
${\cal F}(r)\propto r^{-\vert\alpha(r)-3\vert}$.
If the mass diverges ($2<\alpha(r) <3$, $r\to\infty$),
${\cal F}(r)\sim {\rm const}$.
This is the case of interest here.

\begin{figure}
\plotfiddle{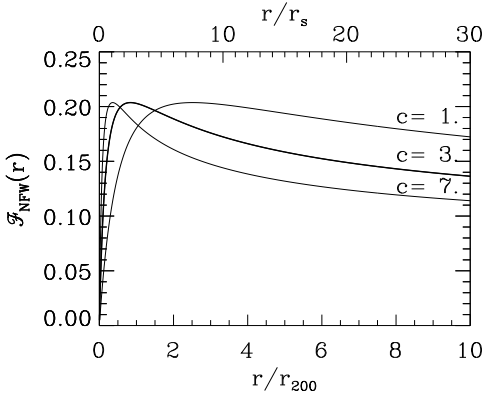}
           {0.25\vsize}              
           {90}                
           {30}                 
           {30}                 
           {100}               
           {-10}                
\caption{Filling function ${\cal F}_{\rm NFW}(r)$ for the NFW density profile (equation 
\ref{ftilde}).
The bold line is ${\cal F}_{\rm NFW}(r/r_s)$. Thin lines are ${\cal F}_{\rm NFW}(r)$ for
different values of the concentration parameter $c=r_{200}/r_s$.}
\label{fig:nfw}
\end{figure}

$N$-body simulations of hierarchical clustering scenarios suggest a
universal density profile for dark matter halos 
(\pcite{NFW}, NFW hereafter).
This profile yields a diverging mass and a convergent gravitational potential at the same time, as
required by the argument above. However, in this case,
$\alpha(r)=3$ when $r\to\infty$, implying $M(<r)\propto \ln(r)$ and
\begin{equation}
{\cal F}_{\rm NFW}(r)  = {1\over 2}{(r/r_s)^2\over(1+r/r_s)^2}
{1\over \ln (1+r/r_s)}
\label {ftilde}
\end{equation}
where $r_s$ is a scale length defined in terms of the halo concentration $c=r_{200}/r_s$, and
$r_{200}$ is the radius of the sphere whose average mass density is 200 times
the critical density. 
${\cal F}_{\rm NFW}(r)$ is not constant in hierarchical
clustering scenarios (see Figure \ref{fig:nfw}). 
However, ${\cal F}_{\rm NFW}(r)$ decreases by $\sim 20$\% at most in the
radius range $r\in[1,10]r_{200}$ for $c\in[1,7]$. This range of concentration $c$ is typical for 
massive halos in Cold Dark Matter (CDM) models; 
in fact, in our $N$-body models (Section \ref{sec:nbody}),
90\% of halos more massive than $10^{14} h^{-1}M_\odot$ have $c$ in the range
$[1.6,5.2]$ with median $c=3.7$.

To see that ${\cal F}_\beta(r)={\cal F}(r)g(\beta)$ is also 
a slowly varying function of $r$,  we compute
the functions $\beta(r)$ and $g(\beta)$ for massive clusters in our $N$-body
simulations (Section \ref{sec:nbody}). Because these two functions do
not show strong variations with $r$, the assumption of a slowly
varying ${\cal F}_\beta(r)$ appears reasonable.
We will specify the value of the constant ${\cal F}_\beta$ in Section \ref{sec:result}.

Finally, it is important to note that
the identification of the amplitude  ${\cal A}(r)$ of the caustics 
with the gravitational potential $[\phi_\beta(r)]^{1/2}$ 
at {\it any} radius $r$ is {\it independent} of the
density profile and the dynamical state of the system, provided that random motions
contribute significantly to the velocity field and the spherical
assumption is a good approximation. On the other hand, the measurement of  ${\cal A}(r)$ yields
the mass profile $M(<r)$ (equation \ref{M2})  only
when the density profile is proportional to $r^{-\alpha(r)}$,
with $2<\alpha(r)\le 3$ when $r\to\infty$.


\section{MEASURING ${\cal A}(\lowercase{r})$}\label{sec:A}

We now outline an operational method for locating the caustics
in the redshift diagram of real clusters. Compiling a redshift 
diagram requires knowledge of the cluster centre and its
radial velocity. Appendix \ref{app:centre} describes how to determine
the cluster centre when only the celestial coordinates of the galaxies and their
redshifts are available.

In order to determine the amplitude ${\cal A}(r)$ of the caustics at
fixed $r$, consider the two-dimensional density distribution function $f_q(r,v)drdv$,
namely the number of galaxies with projected separation in the interval
$(r,r+dr)$ and line-of-sight velocity in the interval $(v,v+dv)$.
In the spherical infall model, at fixed $r$, $f_q(r,v)$ reaches infinity
at the two caustic locations.  Random motions, however, wash these two spikes out and $f_q(r,v)$ has
a maximum close to the cluster redshift; in the absence of massive substructure and
foreground and background galaxies, and in the absence of escaping galaxies, 
the solutions of the equation $f_q(r,v)=0$ would determine
the amplitude  ${\cal A}(r)$. However, this situation never occurs because,
even for isolated systems, escaping galaxies can always be present.
Thus, $f_q(r,v)$ always becomes zero outside the actual location
of the caustics.

We therefore need a recipe for choosing a threshold $\kappa$ such that
the equation $f_q(r,v)=\kappa$ determines the amplitude ${\cal A}(r)$.
Moreover, we have to face the non-trivial issue of 
accurately estimating $f_q(r,v)$ where it is close to zero.
It is clear that the this task is not easy, because
sparse sampling leads to an underestimate of $f_q(r,v)$ and
the presence of foreground and background galaxies leads to an overestimate of $f_q(r,v)$ at
the caustic location.
We describe the determination of $f_q(r,v)$ in Section \ref{sec:fa}. In Section
\ref{sec:kappa} we describe the choice of $\kappa$.

\subsection{Estimating the redshift diagram density distribution}\label{sec:fa}

Consider $N$ galaxies with
coordinates ${\bf x}=(r,v)$, where $r$ and $v$ are conveniently rescaled 
(we come to this issue later). We
use an adaptive kernel method (\pcite{silverman86}; see also \pcite{Pisani93}; \pcite{Pisani96}) 
to estimate the density distribution
of galaxies within the redshift diagram  
\begin{equation}
f_q({\bf x})={1\over N}\sum_{i=1}^{N} {1\over h_i^2} K\left({\bf x}-{\bf x}_i\over h_i\right)
\label {MM1}
\end{equation}  
where 
\begin{equation}
K({\bf t})=\cases { 4\pi^{-1} (1- t^2)^3 & $ t <1$\cr
	0 & otherwise\cr}
\label {MM2}
\end{equation}
and $h_i=h_ch_{\rm opt}\lambda_i$ is a local smoothing length depending on the local density.
The optimal smoothing length is
\begin{equation}
h_{\rm opt} = {3.12 \over N^{1/6}} \left(\sigma_r^2+\sigma_v^2\over 2\right)^{1/2} 
\label {MM3}
\end{equation}
where $\sigma_r$ and $\sigma_v$ are the marginal standard deviations of the galaxy
coordinates.
The local smoothing factor is $\lambda_i=[\gamma/f_1({\bf x}_i)]^{1/2}$ where  
$f_1$ is equation (\ref{MM1}) where $h_c=\lambda_i=1$ for any $i$, 
and $\log\gamma=\sum_i\log[f_1({\bf x}_i)]/N$.

The average degree of smoothing is controlled by $h_c$. The optimal $h_c$ should 
minimize the integrated square error between the estimator $f_q({\bf x})$ and
the true (unknown) density $f({\bf x})$, $\epsilon(h_c)=\int[f_q({\bf x})-f({\bf x})]^2 d^2{\bf x}$.
However, it is easy to show that minimizing $\epsilon(h_c)$ is equivalent to 
minimizing 
\begin{equation}
M_0(h_c) = \int f_q^2({\bf x})d^2{\bf x} - {2\over N} \sum_{i=1}^N f_q^{(-i)}({\bf x}_i)
\label{MM4}
\end{equation}
where $f_q^{(-i)}({\bf x}_i)$ is the density estimated at ${\bf x}_i$ using all
the data points except ${\bf x}_i$ (\pcite{silverman86}). 
Therefore, $M_0(h_c)$ can be estimated with the data alone and we do not need to assume
any form for the true density $f({\bf x})$.

We now address the question of rescaling $r$ and $v$, such
that we can use spherical smoothing windows of size $h_i$.
The adaptive kernel method is designed to estimate density
distributions of random variables. If this were the case, we could
apply the whitening transformation, namely
we would linearly transform the data to have a unit covariance
matrix (\pcite{Fukunaga90}). Unfortunately, ${\bf x}=(r,v)$ is not a random variable and 
the elements of the covariance  matrix
depend on the limits imposed {\it a priori} on the redshift diagram.

We therefore have to rescale
$r$ and $v$ in a sensible way. Let us use the Hubble constant to
have $r$ and $v$ in the same units. It is obvious that, for example,  $100$ km s$^{-1}$ in
the $v$ direction shold not have the same weight as $1 h^{-1}\ {\rm Mpc}=100$ km s$^{-1}$ in 
the $r$ direction. We thus rescale $r$ and $v$ such that 
the ratio $h_v/h_r$ of the smoothing window sizes along $v$ and
$r$ respectively takes a chosen value $q$.
Galaxy redshifts have typical uncertainties of $50$ km s$^{-1}$.
Positions of galaxies within nearby clusters have uncertainties of $\approx0.02 h^{-1}$ Mpc.
We thus set $q=25$ and keep this value fixed hereafter. Note however that
different values of $q$ in the range $[10,50]$ have little effects on the results.
  

\subsection{Choosing the threshold}\label{sec:kappa}

We determine the amplitude ${\cal A}(r)$ at fixed $r$
by finding the solutions of the equation $f_q(r,v)=\kappa$. Specifically, the first upper and
lower solutions $v_{\rm u}$ and $v_{\rm d}$, away from the
maximum of $f_q(r,v)$ closest to $v=0$, determine the amplitude
${\cal A}(r)=\min\{\vert v_{\rm u}\vert,\vert v_{\rm d}\vert\}$. 
Note that the prescription ${\cal A}(r)=\min\{\vert v_{\rm u}\vert,\vert v_{\rm d}\vert\}$ 
is equivalent to  ${\cal A}(r)=(v_{\rm u}-v_{\rm d})/2$
 for an isolated spherically symmetric system. However, our prescription is more
robust than ${\cal A}(r)=(v_{\rm u}-v_{\rm d})/2$ 
against interloper contamination and the presence of massive substructure.

It is clear that, although we have determined $f_q(r,v)$ uniquely (except for the choice of $q$),
there are an infinite number of thresholds $\kappa$ we can use to determine ${\cal A}(r)$.
It seems reasonable to assume that, at least in the central region, the
cluster has attained virial stability. Therefore, in this region,
the equation $\langle v_{\rm esc}^2\rangle_R = 4\langle v^2\rangle_R$ must hold, where
now the angular brackets indicate an average over the whole sphere of radius $R$,
and velocities are three-dimensional. In the data
we only have one-dimensional information, and we thus further 
assume that, if the velocity field is roughly isotropic in the
central region, our expression also holds when 
$v$ is the galaxy line-of-sight velocity and $\langle v_{\rm esc}^2\rangle_{\kappa,R}=
\int_0^R{\cal A}^2(r)\varphi(r)dr/\int_0^R\varphi(r)dr$, where $\varphi(r)=\int f_q(r,v)dv$;
$\langle v_{\rm esc}^2\rangle_{\kappa,R}$ is the
only $\kappa$-dependent quantity. We can choose $\kappa$ by
minimizing the function 
\begin{equation}
S(\kappa,R)=\vert \langle v_{\rm esc}^2\rangle_{\kappa,R} -
4\langle v^2\rangle_R\vert^2. 
\label{eq:S}
\end{equation}
In Appendix \ref{app:centre} we describe a procedure for locating the cluster centre
from a list of galaxy positions.
This procedure also identifies the cluster members. In the following,
we define $R$ as the mean projected distance of the members
from the cluster centre, and $\langle v^2\rangle_R$ the 
one-dimensional velocity dispersion of the cluster members.

The amplitude ${\cal A}(r)$ is obviously sensitive to the value of
$\kappa$. If $f_q(r,v)$ has a rather shallow gradient towards the minimum around the real
location of the caustics, a slightly incorrect estimate of $\langle v^2\rangle_R$  
can lead to particularly inaccurate estimates of ${\cal A}(r)$. Thus, in these cases, 
minimizing $S(\kappa,R)$ will only give a starting value for the choice of $\kappa$.
The final value of $\kappa$ can be chosen {\it by hand} and the location
of the caustics will unfortunately be subjective. 
This freedom is unsatisfactory. On the other hand, $\kappa$ remains
a useful {\it one-parameter} tool which quantifies the complexity of the infall
region of individual clusters and the contamination of their
redshift diagrams by background and foreground galaxies. 
In any case, this subjective
tuning turns out to be necessary in a few cases only, and mainly
for one of the two cosmological models we investigate here, which yields
the poorest fit to observation. This result suggests that the minimization
of $S(\kappa,R)$ is likely to be a robust procedure for real clusters, as the 
Coma cluster has indeed already shown (\pcite{geller99a}).

To control the contamination by background and foreground galaxies efficiently, 
we need a final step. Consider the logarithmic 
derivative $2d\ln{\cal A}/d\ln r=d\ln\phi_\beta/d\ln r$;
we have $d\ln\phi_\beta/d\ln r=d\ln\vert\phi\vert/d\ln r-d\ln g/d\ln r$.
For any realistic system $d\ln\vert\phi\vert/d\ln r\la 0$. Our $N$-body
simulations show that $d\ln g/d\ln r \ga -1/2$. Thus,
we can safely claim that $d\ln\phi_\beta/d\ln r \la 1/2$ should hold for any $r$.
To be conservative, we therefore accept only values of ${\cal A}(r)$
which yield $d\ln{\cal A}/d\ln r<1$; otherwise we impose a new value of ${\cal A}(r)$
which yields $2d\ln{\cal A}/d\ln r=1/2$.

\begin{table}
\caption{GIF Simulations}\label{tab:gif}
\begin{tabular}{rrrrrrr}
model & $\Omega_0$ & $\Omega_\Lambda$ & $h$   & $\sigma_8$ & $m_p$ & $L$ \\
$\Lambda$CDM & 0.3 & 0.7 & 0.7 &  0.90 & 1.4 & 141 \\
$\tau$CDM & 1.0 & 0.0 & 0.5 &  0.60 & 1.0 & 85 \\
\end{tabular}

\medskip
{Parameters of the two GIF simulations (\pcite{kauffmann98}) used in this paper.
The Hubble constant $h$, the particle mass $m_p$, and
the comoving size $L$ of the simulation box are in units of $H_0=100$ km s$^{-1}$ Mpc$^{-1}$,
$10^{10}$ $h^{-1}$ $M_\odot$, and $h^{-1}$ Mpc, respectively.}
\end{table}

\begin{figure*}
\plotfiddle{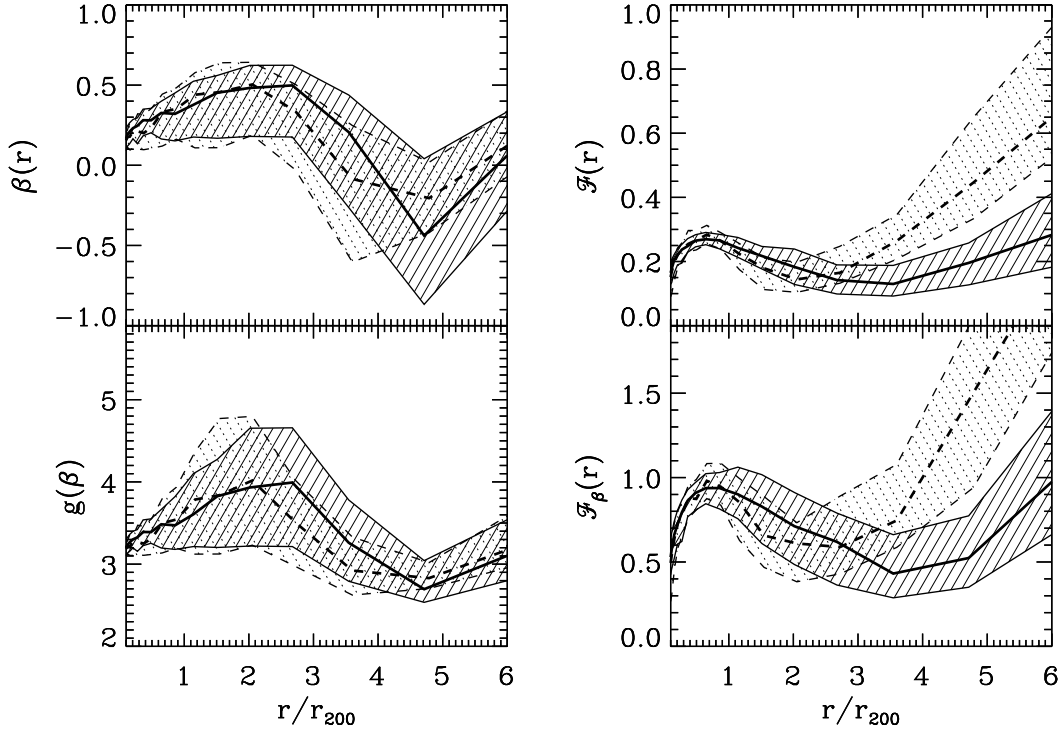}
           {0.45\vsize}              
           {90}                
           {60}                 
           {60}                 
           {230}               
           {-30}                
\caption{Profiles of $\beta(r)$, $g(\beta)$, ${\cal F}(r)$, and ${\cal F}_\beta(r)$
for the clusters with $M_{200}\ge 10^{14}h^{-1} M_\odot$ in the $\tau$CDM (dashed lines)
and $\Lambda$CDM (solid lines) models.
Bold lines are the median profiles. Shaded areas show the interquartile range of the profiles.}
\label{fig:fb}
\end{figure*}

\section{NUMERICAL MODELS}\label{sec:nbody}
 
In Section \ref{sec:result} we apply our mass estimation method to  
dissipationless $N$-body simulations where we form and evolve galaxies 
within their dark matter halos with phenomenological recipes.
Here, we briefly describe our cosmological simulations. 

\subsection{The GIF simulations}

We use two $N$-body simulations from the GIF project
(\pcite{kauffmann98}). These simulations use $256^3$ dark matter particles to model the evolution, 
from redshift $z=50$ to the present, of the dark matter density perturbations 
of a CDM universe with an initial power spectrum 
\begin{equation}
P(k) \propto k \{1+[6.4k/h\Gamma+(3k/h\Gamma)^{3/2} +
(1.7k/h\Gamma)^2]^{1.13}\}^{-2/1.13}
\end{equation}
where $k$ is in units of $h$ Mpc$^{-1}$, and $\Gamma=0.21$ 
is a shape parameter (\pcite{efstathiou92}).
The normalization is fixed by the ratio $\sigma_8^2$ of the variances of the mass 
and galaxy fluctuations within randomly placed spheres of radius $8h^{-1}$ Mpc. The GIF models
are normalized to give the correct abundance of rich galaxy clusters at the present time.   
The simulations were run with {\it Hydra} (\pcite{pearce98}), the parallel version of the 
AP$^3$M code (\pcite{couchman91}; \pcite{couchman95}), kindly provided by
the Virgo supercomputing consortium (\pcite{jenkins97}). 
The two models we consider here are a flat model with or without a cosmological constant: 
$\Omega_0=0.3$ and $\Omega_\Lambda=0.7$ ($\Lambda$CDM) and $\Omega_0=1$ ($\tau$CDM).
Table \ref{tab:gif} summarizes the parameters of these models.

\scite{kauffmann98} 
combine these $N$-body simulations with semi-analytic modelling 
to form and evolve galaxies within dark matter halos. 
They provide a detailed description of this procedure and compare the observable properties
of the simulated galaxy catalogues with the real Universe in a series of papers
(\pcite{kauffmann98}; \pcite{kauffmann99}; \pcite{Diaferio99}).
The relevant physical processes for galaxy formation include
 gas cooling, star formation, supernova feedback, stellar evolution, and
merging of galaxies. Previous
attempts to compare such dissipationless simulations with the real galaxy
distribution were based on some high-peak statistical model for galaxy formation or on 
{\it ad hoc} assumptions 
about the mass-to-light ratio of dark matter halos (e.g. \pcite{frederic95}; \pcite{nolthenius97}). 
Alternatively,  we could consider a full $N$-body/hydrodynamic
simulation. However, the current state-of-the-art for such simulations 
can model only a small volume of the Universe, otherwise
their resolution limit is larger than the size of galaxies and their halos
(e.g. \pcite{frenk96}; \pcite{navarro97a}; \pcite{weinberg97}; \pcite{Cen99}; \pcite{Blanton99};
\pcite{pearce99}).

The GIF  simulations have a resolution limit of $30 h^{-1}$ kpc and
 are suitable for investigating the reliability of
our mass estimation method. We can in fact construct
mock redshift diagrams with {\it galaxies}, along with their observable
properties, rather than with dark matter particles alone. 
Note that, in these models, galaxy samples 
show only a weak velocity bias, independently of galaxy luminosities, as expected if gravity
is the driving force on scales $\ga 1h^{-1}$ Mpc.

As discussed in \acite{kauffmann98}, 
the free parameters entering the galaxy formation recipe 
have substantial effects on galaxy properties.
Here, for consistency, we consider the ``fiducial'' models 
which yield reasonable fits to many but not all the observed properties of galaxies;
in particular the model luminosity function is a rather poor fit to observation. 
For the purpose of this paper, it is worth noting 
that the $\tau$CDM model has a luminosity density in the $B$-band which is 
a factor $\sim 2$ {\it larger} than in real galaxy samples, whereas
the $\Lambda$CDM model has a luminosity density a factor $\sim 2$ {\it smaller}. 
In order to quantify the effect of the luminosity function 
on the derived properties of the galaxy distribution, \scite{Diaferio99}
assign new luminosities to the model galaxies, while preserving
their luminosity rank in such a way as to reproduce 
the luminosity function of the Center for Astrophysics (CfA)
redshift survey (\pcite{marzke94}) exactly. We therefore have two distinct sets of simulated galaxy  
catalogues: (1) the SALF catalogues where galaxies have luminosities derived from the semi-analyitic
modelling, and (2) the CfALF catalogues where galaxies have luminosities imposed according
to the CfA luminosity function.

\subsection{Profiles of dark matter halos}

Before testing the mass estimation method, we address the 
robustness of the assumptions described in Section \ref{sec:mass},
namely whether ${\cal F}_\beta(r)$ is a slowly varying function of $r$
in these CDM models.

We identify dark matter halos using a friends-of-friends group finder which
links particles closer than 0.2 times the interparticle separation.
We take the position of the most bound particle as the centre of the halo.
Here, we consider only dark matter halos with $M_{200}\ge 10^{14}h^{-1}M_\odot$, where
$M_{200}$ is the mass within $r_{200}$.
Less massive halos are less interesting for our purpose, because our
mass estimation method, which requires the redshifts of a few hundred galaxies, 
will be applicable to massive clusters only.

Figure \ref{fig:fb} shows the median profiles of the relevant quantities of
the halos in both models.
Shaded areas indicate the interquartile range of the profiles. 
The velocity anisotropy parameter $\beta(r)$ 
increases from $\sim 0.2$ to $\sim 0.5-0.6$ between $r=0.1 r_{200}$ and $r\sim 3r_{200}$,
as radial motions become predominant. Remarkably, this result agrees with the
value $\beta\sim 0.5$ estimated by \scite{biviano97} for the emission-line galaxies 
in the ESO Nearby Abell Cluster Survey.
At the turnaround radius $r \approx 4-5r_{200}$, $\langle v_r^2\rangle$ 
decreases faster than its non-radial counterparts $\langle v_\phi^2\rangle$ 
and $\langle v_\theta^2\rangle$; 
thus, $\beta(r)$ drops to negative values. 
Note that $\beta(r)$ is sensibly smaller than one everywhere,
indicating that random motions always
contribute significantly to the velocity field, as expected in hierarchical clustering
scenarios.
The function $g(\beta)$ we introduced in Section \ref{sec:rd} 
behaves as $\beta(r)$. More importantly, $g(\beta)$ varies with $r$ slowly,
 increasing by 30-50\% at most between $r=0.1 r_{200}$ and $r\sim 2-3r_{200}$. 
At radii where the Hubble flow dominates, $\beta(r)\to 1$ and $g(\beta)\to \infty$.

To compute the function ${\cal F}(r)$ introduced in Section \ref{sec:mass}, 
we need to compute the gravitational potential $\phi(r)$ generated by the halo.
We could compute $\phi(r)$ with equation (\ref{M4}) where we replace the upper
limit of integration with a maximum radius. However, because the halo density
profiles are well approximated by the NFW formula, it is more efficient to 
obtain a fit to the density profile and compute $\phi(r)$ analytically.
This procedure bypasses the problem of subtracting the background at large radii.
Figure \ref{fig:fb} shows that ${\cal F}(r)$ is slowly varying, as expected.  
Note that at radii $r \ga 3-4r_{200}$, the derivative of ${\cal F}(r)$ is positive.
In fact, the density profiles do not 
decrease as $r^{-3}$ as predicted by the NFW profile, but rather 
decrease as $r^{-\alpha}$, with $\alpha\sim 1-2$,
because they must approach the background density. 
Therefore ${\cal F}(r)\sim -\rho r^2/\phi$ increases as $r^{3-\alpha}/\ln r$.
Note that the statistical spread in the ${\cal F}(r)$ profiles 
accounts for different concentration parameters $c$.
As expected,  ${\cal F}_\beta(r)$  has a trend similar to the trend of 
${\cal F}(r)$,  because $g(\beta)$ does not vary strongly.
Variations of $g(\beta)$ only make the statistical spread of ${\cal F}_\beta(r)$ larger.

Note that the difference between the $\tau$CDM and the $\Lambda$CDM models
for each of the four profiles
is mainly due to the different physical location of $r_{200}$ in 
the clusters of  the two models: $r_{200}$ identifies a sphere with an average density, 
in units of the {\it background} density, which is larger in the low density universe 
than in the high-density one.
 
In conclusion, ${\cal F}_\beta(r)$ is not constant in these models,
as assumed in equation (\ref{M2}), but it varies by a factor of two at most within
$r\sim 3-4r_{200}$. Despite these variations, equation (\ref{M2}) remains a
useful recipe, as we show in the next section. 

\section{GALAXY CLUSTERS}\label{sec:result}

We now apply our caustic location method and mass estimation procedure to simulated clusters.
To construct the redshift diagram, 
we use the centre of the dark matter halo and include
galaxies brighter than $M_B=-18.5+5{\rm log}h$. This magnitude
limit corresponds to an apparent magnitude of $m_{\rm lim}=15.7$
at the distance of Coma ($R=7000$ km s$^{-1}$) and is one magnitude brighter than the limit
of completeness of our galaxy catalogues (\pcite{kauffmann98}). 

\begin{figure*}
\plotfiddle{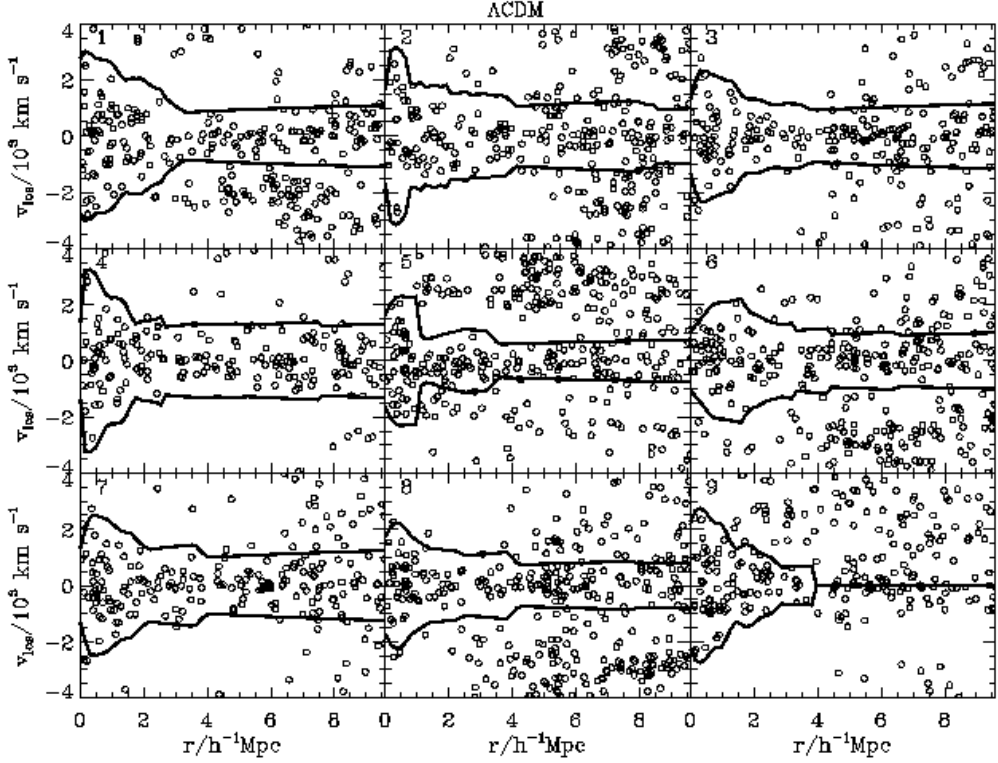}
           {0.80\vsize}              
           {90}                
           {60}                 
           {60}                 
           {230}               
           {-30}                
\caption{Redshift diagrams of a cluster in the $\Lambda$CDM model viewed
along nine different lines of sight. The circles show galaxies
brighter than $M_B=-18.5+5{\rm log}h$. The bold lines show the identified
caustic locations. When the contamination by foreground 
and/or background galaxies is too heavy for a reliable location
of the caustics the method does not proceed further.
This situation is represented by merging the caustics into a single horizontal line 
at the cluster redshift (see right-bottom panel).} 
\label{fig:lambda_rd}
\end{figure*}

\begin{figure*}
\plotfiddle{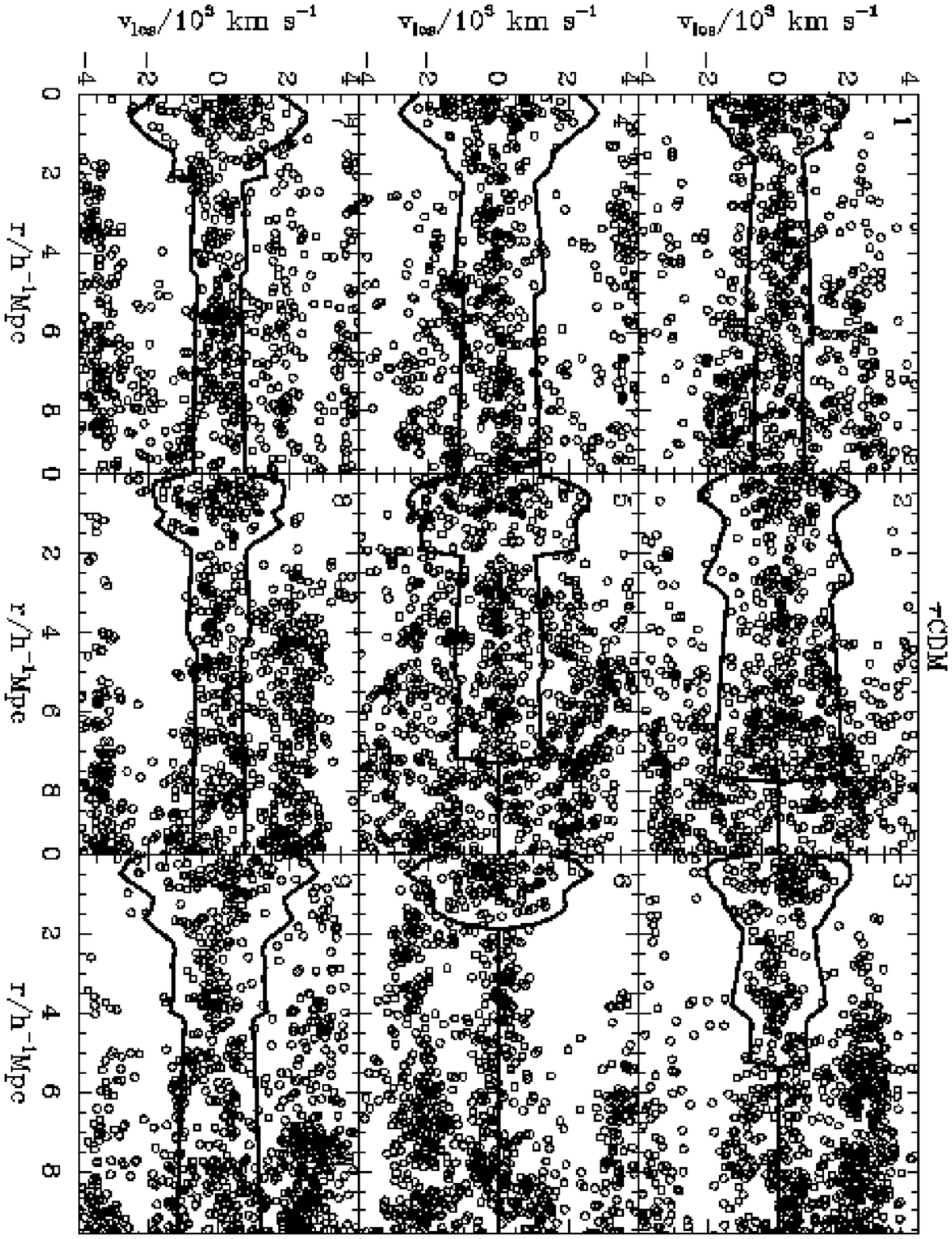}
           {0.80\vsize}              
           {90}                
           {60}                 
           {60}                 
           {230}               
           {-30}                
\caption{Same as Fig. \ref{fig:lambda_rd} for a cluster in the $\tau$CDM model.}
\label{fig:tcdm_rd}
\end{figure*}

We also simulate observations of nearby clusters, by randomly choosing 
the observer's location on a sphere of radius $R=7000$ km s$^{-1}$ centered on
the centre of the dark matter halo. We then compile a list of galaxies brighter
than $m_{\rm lim}=15.7$ with
coordinates $(\alpha,\delta,cz)$ within 10$^{\rm o}$ from the halo centre; 
from this list, we locate the cluster centre with the method described
in Appendix \ref{app:centre}. This method usually locates the cluster centre 
accurately; both this procedure and the one which uses
the centre of the dark matter halo directly, yield similar results. 
Below, we show results where we use the centre  of  the dark matter halo.

\subsection{Caustic Location}

Fig. \ref{fig:lambda_rd} shows the redshift diagrams of a typical cluster in the $\Lambda$CDM model.
The nine panels correspond to nine different lines of sight.
The galaxies tend to populate a defined region in the redshift diagram.
The contrast is quite  evident between the galaxies within and outside this
region. In most cases, our method seems to locate the caustics, i.e. the borders of this
region, correctly.

As discussed in \scite{Diaferio99}, despite the fact that mock redshift surveys extracted from 
both the $\tau$CDM and the $\Lambda$CDM models do not show structures
as sharply defined as in real surveys, the two models
still show substantial differences: the $\Lambda$CDM model
shows voids and filaments larger than the $\tau$CDM model and yields
a better qualitative fit to the real Universe (see \pcite{schmalzing99} for
a quantitative analysis). These differences
also are quite apparent in redshift diagrams of clusters 
(Figs. \ref{fig:lambda_rd} and \ref{fig:tcdm_rd}), where the density contrast between interlopers
and galaxies falling onto the cluster is less evident in the $\tau$CDM than 
in the $\Lambda$CDM model. 
Nevertheless, in $\tau$CDM clusters, our method still 
locates the caustics reasonably well in most cases (Fig. \ref{fig:tcdm_rd}). 

The different redshift diagrams yielded by the two cosmological 
models are related to the underlying mass distribution
rather than to the galaxy luminosity function.  
Using the CfALF catalogue decreases the
number of galaxies in the redshift diagrams of $\tau$CDM clusters
but does not improve the appearence of the caustics. In fact, 
in this high density universe, clusters are still accreting mass  
at a large rate at redshift $z=0$. On the other hand, in 
the $\Lambda$CDM, where using the CfALF catalogue increases
the number of galaxies in the redshift diagrams, the accretion
rate is substantially smaller and the caustics tend to appear more clearly,
as observed in real clusters (\pcite{geller99a}).  

In principle, this redshift diagram morphology could be used to distinguish
between high and low density universes.
For example, we could quantify this difference by using the distribution of the 
ratio between the number of galaxies lying 
within the caustics and the total number of galaxies within the redshift diagram.
This ratio, of course, depends on the cluster
and on the line of sight, but the moments of the  distribution should be 
sensitive to the cosmological model.
For example, these ratios for the different lines of sight of 
the $\Lambda$CDM cluster (Fig. \ref{fig:lambda_rd}) 
are in the range $[0.32,0.88]$ 
with median 0.53 (panel 2) and are considerably larger than the corresponding
ratios of the $\tau$CDM cluster (Fig. \ref{fig:tcdm_rd}): 
$[0.11,0.55]$ with median 0.39 (panel 7). These distributions appear
to be robust against variations of the magnitude limit of the galaxy sample. 
The feasibility of this kind of statistical
tests will be investigated in future work.

In panels 2, 5, and 9 of Fig. \ref{fig:tcdm_rd} the
caustics seem to be outside the location we would guess subjectively, because
of the heavy contamination by foreground and background galaxies.  
Our intuition is confirmed by 
comparing the amplitude ${\cal A}(r)$ of the caustics
with the function $[\phi_\beta(r)]^{1/2}$
computed from the full six-dimensional phase space information. The upper right
panel of Fig. \ref{fig:comb_prof} shows all nine estimated profiles
for the $\tau$CDM cluster: the 
ones which tend to overestimate ${\cal A}(r)$ come from the three errant panels.
In the other six redshift diagrams and in all the diagrams
of the $\Lambda$CDM cluster the agreement is rather good. 

\begin{figure*}
\plotfiddle{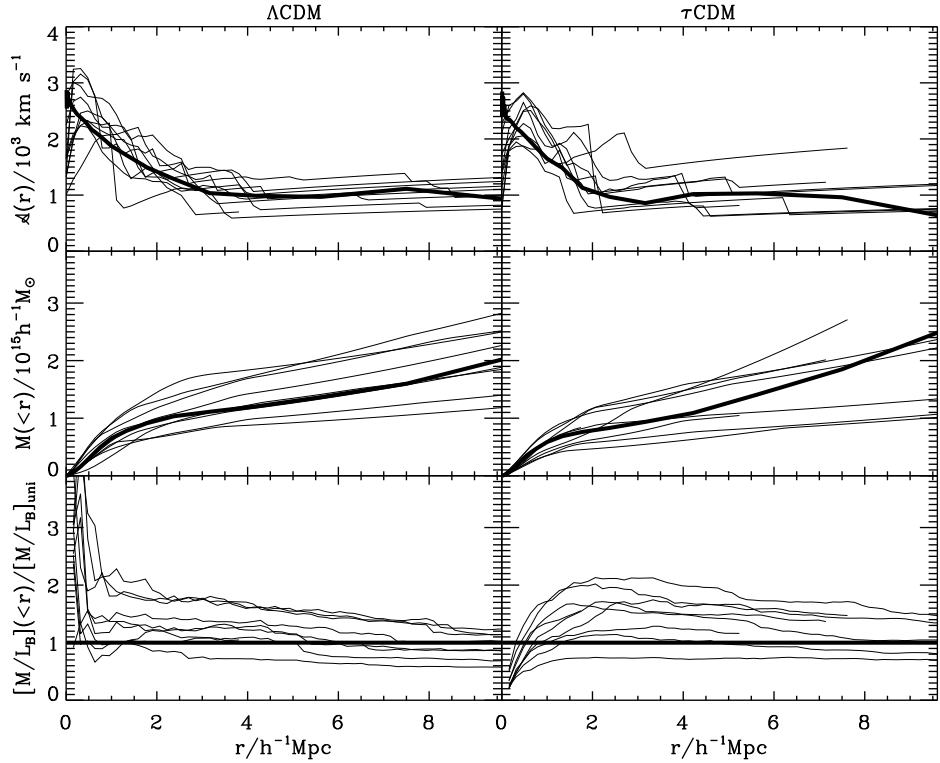}
           {0.90\vsize}              
           {90}                
           {60}                 
           {60}                 
           {230}               
           {-30}                
\caption{Profiles of the two clusters shown in Figs. \ref{fig:lambda_rd} and
\ref{fig:tcdm_rd}.  The thin lines are the profiles measured from the redshift diagrams.
The bold lines in the four upper panels are the true profiles.}
\label{fig:comb_prof}
\end{figure*}

It is worthwhile to note, at this point, that our caustic location procedure is essentially
an interloper-removal procedure which uses the combined information on 
the galaxy position and line-of-sight velocity, unlike standard 3-sigma clipping
procedures which ignore the position of galaxies (e.g. \pcite{yahil77}; \pcite{perea90}). 
Our procedure enables us to estimate the mass
profile of clusters as a by-product, on the assumption that
clusters form through hierarchical clustering.\footnote{An 
interloper-removal procedure, which also
uses both the position and the velocity of galaxies,
was introduced by \scite{denhartog96}.
Their iterative procedure is effective in determining the cluster membership, but it
cannot be used to estimate the cluster mass. In fact, this  procedure 
estimates the profile of the maximum line-of-sight velocity allowed for a cluster member,
namely a ``caustic'', {\it only after} estimating the mass profile. Moreover,
the estimate of the mass profile relies on
the virial theorem all the way out to the infall region.}

Let us finally emphasize that our procedure does not remove
all the interlopers: in fact, any interloper-removal procedure, which
does not separate the galaxy redshift in its Hubble flow and peculiar
velocity components, is obviously unable to identify interlopers lying
within the caustics in redshift space.

\subsection{Mass Profiles}

To compute the mass profile from the caustic amplitude, 
we set ${\cal F}_\beta=1/2$ in equation (\ref{M2}),
as first suggested by DG, although Fig. \ref{fig:fb} shows that
the mean value of ${\cal F}_\beta$ may be slightly larger; moreover, we set
$r_0=0$. We thus have 
\begin{equation}
GM_{\rm est}(<r) = {1\over 2} \int_0^r {\cal A}^2(x) dx.
\label {subm1}
\end{equation}
In writing equation (\ref{M2}) we have assumed that ${\cal F}_\beta(r)$ is
roughly constant at {\it large} $r$. Thus, setting $r_0=0$ is
justified if ${\cal F}_\beta(r)$ is also roughly constant at {\it small} $r$
and takes the same values as at a large $r$.
In fact, Fig. \ref{fig:fb} shows that ${\cal F}_\beta(r)\in [0.5,1.0]$ for
any $r\la 4r_{200}$, and Fig. \ref{fig:comb_prof} shows that the estimate of
$M(<r)$ at small $r$ is as good as at large $r$.

The mass profile is recovered within an uncertainty of 50\% out 
to $\sim 8 h^{-1}$ Mpc for the $\Lambda$CDM cluster (left middle panel
of Fig. \ref{fig:comb_prof}). For the $\tau$CDM cluster the estimated
mass profile is less accurate at those distance, but still within
a factor of two.

Finally, the bottom panels of Fig. \ref{fig:comb_prof} 
show the profiles of  the $B$-band mass-to-light ratio 
in units of the mean ratio for the entire
simulation box. In the $\Lambda$CDM cluster, the mass-to-light ratio 
tends to be larger than the universal value
at very small radii because of the deficiency of blue galaxies in the
central region of clusters
in this cosmology \cite{Diaferio99}. At radii $\sim 4-6h^{-1}$ Mpc 
both models show a mass-to-light ratio consistent
with the universal value. Note that for these massive clusters, on average, the 
mass-to-light ratio within $r_{200}\sim 1.5h^{-1}$ Mpc, computed with the full three-dimensional
information, is close to the universal value (see Fig. 15 of \pcite{kauffmann98}).
This result is consistent with our Fig. \ref{fig:comb_prof} where the
estimation of the mass-to-light ratio at $r\sim 1.5h^{-1}$ Mpc suffers from 
projection effects and errors in the mass estimate.

Real clusters, of course, provide a single redshift diagram. 
The error in the measured value of ${\cal A}(r)$ should depend on the
number of galaxies which contribute to the determination of ${\cal A}(r)$.
We thus assume that the relative error $\delta{\cal A}(r)/{\cal A}(r)=
\kappa/\max\{f_q(r,v)\}$, where the maximum value is found 
along the $v$-axis at fixed $r$. 
We then define the error in the mass profile as
$\delta M_i=\sum_{j=1,i}\vert 2m_j\delta{\cal A}(r_j)/{\cal A}(r_j)\vert$, 
where $m_j$ is the mass of the shell $[r_{j-1},r_j]$ given by equation (\ref{M2})
with ${\cal F}_\beta=1/2$. This recipe yields errors in agreement
with the typical spread due to the projection effects shown in Fig. \ref{fig:comb_prof}.

\section{DISCUSSION}

DG suggested the possibility of using redshift data alone to measure the mass 
of clusters  within $\la 10h^{-1}$ Mpc from their centres. 
This mass, combined with photometric measurements, provides an estimate
of the mass-to-light ratio and therefore the mean mass density
of the Universe, $\Omega_0$, on the assumption that the value obtained
on this relatively large scale is close to the global value.

Here, we describe
an operational procedure which can be applied to redshift diagrams of real clusters
containing a few hundred galaxies with measured redshifts.
We apply this procedure to galaxy clusters simulated from
$N$-body models which include semi-analytic
modelling of galaxy formation. We are thus able to mock observations
of clusters where the luminosity and the formation history of galaxies are included.
We recover the actual cluster mass profile within a factor of
two to several megaparsecs from the cluster center.

For the sake of clarity we summarize here the assumptions and the parameters
entering our mass estimation procedure, giving
in parentheses the section where we discuss the issue extensively. 
We assume that (1) clusters are spherically symmetric (Section \ref{sec:rd}); 
(2) clusters form through hierarchical
clustering: this assumption implies that (i) non-radial motions are an important component of 
the velocity field in the infall region of clusters (Section \ref{sec:rd}),
and (ii)  the filling function ${\cal F}_\beta(r)$, which combines
the cluster  density profile and the anisotropy of the
velocity field, is roughly constant at large distance $r$ from the
cluster centre (Section \ref{sec:mass}); (3) 
substructures in the cluster infall region have mass substantially
smaller than the cluster mass (Section \ref{sec:rd}); in
other words, we cannot apply our mass estimation method when 
a major merging between clusters is taking place.  

The parameters 
governing the location of the caustics and the estimate of the mass profile are: (1) the ratio
$q$ between the sizes of the smoothing window along the line-of-sight
velocity $v$ and the angular separation $r$ (Section \ref{sec:fa}); (2) 
the threshold $\kappa$ for the redshift space distribution function
$f_q(r,v)$ (Section \ref{sec:kappa}); (3) the maximum value
allowed for the logarithmic derivative of the caustic amplitude ${\cal A}$ (Section \ref{sec:kappa}); 
(4) the filling factor ${\cal F}_\beta$ assumed to be constant over
the entire interval of the caustic amplitude integration (Section \ref{sec:mass}, 
equation \ref{M2}; Section \ref{sec:result}, equation \ref{subm1}).

Our procedure is automatic and, despite the four parameters listed above, 
in essence non-parametric.
In fact, $q$ has been kept fixed throughout our analysis and the result are little
affected by its variations. The maximum logarithmic derivative  of ${\cal A}$
also has been kept fixed to a rather large value which is rarely reached.
Finally, the choice ${\cal F}_\beta=1/2$ is a consequence of the assumption 
of the validity of the hierarchical clustering scenario.
The parameter $\kappa$ is determined by an automatic procedure.
However, particularly unfortunate situations may require
a tuning of this parameter to locate the caustics
accurately. The subjective choice of $\kappa$ is
required more often in the $\tau$CDM model than in the $\Lambda$CDM model.
The $\Lambda$CDM model yields redshift diagrams more similar to
those of real clusters, where the density contrast between galaxies
falling into the cluster and unrelated galaxies is more evident.
Therefore, our method is likely to be robust when applied to real clusters
(\pcite{geller99a}).

\scite{vedel98} suggest an 
alternative procedure to extract the information contained in redshift diagrams.
This procedure is based on a maximum likelihood technique and
therefore has the disadvantage of being constrained by
the assumed model for the velocity and density profiles.
In hierachical clustering scenarios, the infall region dynamics
can be very complex and modelling these profiles might turn
out to be a particularly difficult task. This procedure is likely to be  
successful only when we average over many clusters.

At redshift $z\ga 0.2$, methods based on weak gravitational lensing 
(e.g. \pcite{kaiser95}; \pcite{seitz96}; \pcite{squires96}; \pcite{lombardi98})
can also provide a mass estimate in the outer regions 
of clusters. These methods measure all the mass
projected along the line of sight and suffer from systematics due
to contributions from the large scale structure (e.g. \pcite{bartelmann95};
\pcite{reblinsky99}); 
moreover, these methods cannot be applied to nearby
clusters. On the other hand, our approach measures the local mass and,
in principle, is not constrained to any redshift range.

Measurements of galaxy redshifts of several nearby clusters are already
currently available. \scite{geller99a} 
have used this method  to measure
the mass profile of Coma out to $10h^{-1}$ Mpc from the cluster centre. On
smaller scales this profile encouragingly agrees with estimates based
on X-ray observations. Applications to other nearby Abell clusters are currently underway.

\vskip 1truecm
\noindent{\bf ACKNOWLEDGMENTS}

\noindent
I sincerely thank Margaret Geller for stimulating my interest in the dynamics of the infall
region of galaxy clusters. Her inexhaustible enthusiasm 
made this work possible.  I especially thank J\"org Colberg for many
fruitful discussions and competent suggestions concerning non-trivial computer riddles
at an early stage of this project. 
This work also benefited from discussions with 
Matthias Bartelmann, David Chernoff, Bhuvnesh Jain, 
R\"udiger Kneissl, Thomas Loredo, Peter Schneider, 
Ravi Sheth, Bepi Tormen, Roberto Trasarti Battistoni, 
Rien van de Weygaert, Ira Wasserman, Simon White, and Saleem Zaroubi.
I thank an anonymous referee whose relevant suggestions improved the presentation
of my mass estimation technique. The $N$-body simulations were carried
out at the Computer Center of the Max-Planck Society
in Garching and at the EPPC in Edinburgh, as part of the Virgo Consortium project.
During this project, I was a Marie Curie fellow and held the grant ERBFMBICT-960695 
of the Training and Mobility of Researchers
program financed by the European Community. I also acknowledge support from 
an MPA guest post-doctoral fellowship.

\appendix
\section{LOCATING THE CLUSTER CENTRE}\label{app:centre}

Defining the centre of a real cluster is not trivial. Depending on the available data, we can
define the centre as the position of the cD or the D galaxy, or
as the position of the peak of the X-ray emission (see e.g. \pcite{denhartog96}). 
However, these definitions are not
unique; clusters may contain more than one D galaxy or may have multiple
X-ray peaks.

Here, according to the definition of the cluster centre adopted in our $N$-body simulations, 
we wish to locate an observable cluster centre as close as possible 
to the minimum of the gravitational potential well of the cluster.
We therefore locate the centre of the cluster with a two-step
procedure: (1) a hierarchical method identifies galaxies in the
sample which are cluster members; (2) an adaptive kernel method estimates
the cluster member density distribution projected onto the sky. We define
the centre of the cluster as the peak of this
distribution; the cluster centre in redshift space is the median of the
cluster member velocity distribution. 
Note that the main goal of the hierarchical method (step 1) is the identification of
the cluster substructure rather than the cluster centre.  
The determination of the cluster centre with
step 2 is a natural by-product of the substructure analysis.

We identify members of the galaxy cluster with a hierarchical cluster analysis (e.g. 
\pcite{Miyamoto90}).
Cluster analysis classifies $N$ objects using a measure of similarity between any
two objects. Cluster analysis produces a binary tree with similarity decreasing
from the leaves to the root. At any level of the hierarchy the binary tree provides    
a number of distinct groups: two members within the same groups have similarity larger
than two members within two different groups. 

Different definitions of similarity have been applied to galaxy catalogs (\pcite{Materne78};
\pcite{Tully87}; \pcite{Gourgoulhon92}).
Recently, \scite{Serna96} suggested using the galaxy pairwise binding
energy as a measure of similarity: 
\begin{equation}
E_{ij} = -G{m_im_j\over \vert {\bf r}_i-{\bf r}_j\vert} +
{1\over 2} {m_im_j\over m_i+m_j}({\bf v}_i-{\bf v}_j)^2
\label{eq:be}
\end{equation}
where $G$ is the gravitational constant, $m_i$, $m_j$, ${\bf r}_i$, ${\bf r}_j$, 
${\bf v}_i$, ${\bf v}_j$ are the masses, positions and velocities of the two
galaxies. The reliability of the relative binding energy as a measure of similarity
has been questioned by \scite{Gurzadyan98a} who suggest a more physically
motivated method of grouping galaxies. Indeed, 
when applied to observed systems, equation (\ref{eq:be}) has two incoveniences: 
(1) the masses $m_i$, $m_j$ of the two galaxies are unknown; (2) projected information
provides only three out of the six phase-space dimensions. 
By ignoring the unknown components of the position and velocity vectors, we obtain values of the
binding energy $E_{ij}$ which are smaller than the $E_{ij}$'s computed with the full six-dimensional
information; moreover, the rank of $E_{ij}$  of different pairs is not
necessarily preserved. 
Despite these shortcomings, 
\scite{Serna96} tested this method with $N$-body simulations and obtained resuls which are generally
more satisfying than other grouping algorithms. 
Moreover, comparison with our simulations shows that the centre located
with this definition of similarity is reasonably close to the minimum of the
potential well, as we require.

Each galaxy is located by its vector in redshift space ${\bf s}_i=(\alpha_i,\delta_i,cz_i)$,
where $\alpha_i,\delta_i$ are its celestial coordinates and $cz_i$ is its radial velocity.
Each galaxy pair defines
the line of sight vector ${\bf l}=({\bf s}_i+{\bf s}_j)/2$ and the
separation vector ${\bf s}={\bf s}_i-{\bf s}_j$. Thus, we obtain
the line-of-sight component $\pi$ of the velocity difference and
the separation projected onto the sky $r_p$ 
\begin{equation}
\pi = {{\bf s}\cdot{\bf l}\over \vert{\bf l}\vert};
\quad r_p = ({\bf s}^2 - \pi^2)^{1/2}.
\label{eq:proj}
\end{equation}
In redshift space, equation (\ref{eq:be}) becomes
\begin{equation}
E_{ij} = -G{m_im_j\over r_p} +
{1\over 2} {m_im_j\over m_i+m_j}\pi^2.
\label{eq:bep}
\end{equation}

To build the binary tree we proceed as follows:
\begin{enumerate}
\item each galaxy is a group $G_\nu$; 
\item we compute the similarity between two groups $G_\mu$, $G_\nu$ with the single linkage 
method: $E_{\mu\nu}=\min\{E_{ij}\}$ where $E_{ij}$ is the similarity between the member
$i\in G_\mu$ and the member $j\in G_\nu$; 
\item we replace the two groups with the largest similarity (smallest binding energy $E_{ij}$) 
with a group $G_\kappa$. The number of independent groups is decreased by one;
\item the procedure is repeated from (ii) until we are left with only one independent group.
\end{enumerate}

If we want a catalogue of disjoint groups 
we need to choose a threshold where we cut the tree.
This threshold is somewhat arbitrary (\pcite{Materne78}). 
Note that we use the binding energy as a similarity parameter, 
so we cannot use thresholds based on the luminosity density (\pcite{Tully87}).
The arbitrariness of the threshold is
intrinsic to the long range of the force of gravity: clumps are less and less bound as we climb
up the tree (i.e. we go towards the root) but they are never completely independent.

In order to identify the final threshold which determines the members of the main
cluster, we first compile a list of candidate thresholds; 
from this list  we then identify the final threshold.

We adopt the following procedure to compile the list of candidate thresholds. This procedure 
does not depend on the actual value of the binding energy
computed with equation (\ref{eq:be}): therefore the value adopted for the
galaxy mass is irrelevant.

We walk through the tree along the main branch. At each step the main 
branch having $n_d$ descendents (leaves) splits into two branches. 
When both branches contain a substantial fraction of the parent descendents
we label the parent of the largest branch as a threshold. In other words, 
whenever the smallest branch contains a number of descendents $n_s>fn_d$,
where $f$ is a free parameter (we use $f=0.1$), 
the level  of the largest branch is a threshold. Thus, we obtain a list
of thresholds where the main branch looses a substantial fraction of its descendents.

We now need to identify the final threshold. If the cluster is not completely relaxed, 
galaxies within the dark matter halo of the cluster may be distributed within 
subclumps. $N$-body simulations (e.g. \pcite{Tormen98}) suggest that the dark matter halos of 
these subclumps are easily disrupted within the main halo, thus 
galaxies roughly feel the same potential well, regardless of their parent subclump.
Therefore, we do not expect the velocity dispersion $\sigma$ to vary substantially when different
subclumps are included or excluded from the computation of $\sigma$. We expect a substantial 
increase of $\sigma$ when we include galaxies outside the main halo or a decrease of $\sigma$ 
when we consider the very central galaxies of a given subclump.  

\begin{figure}
\plotfiddle{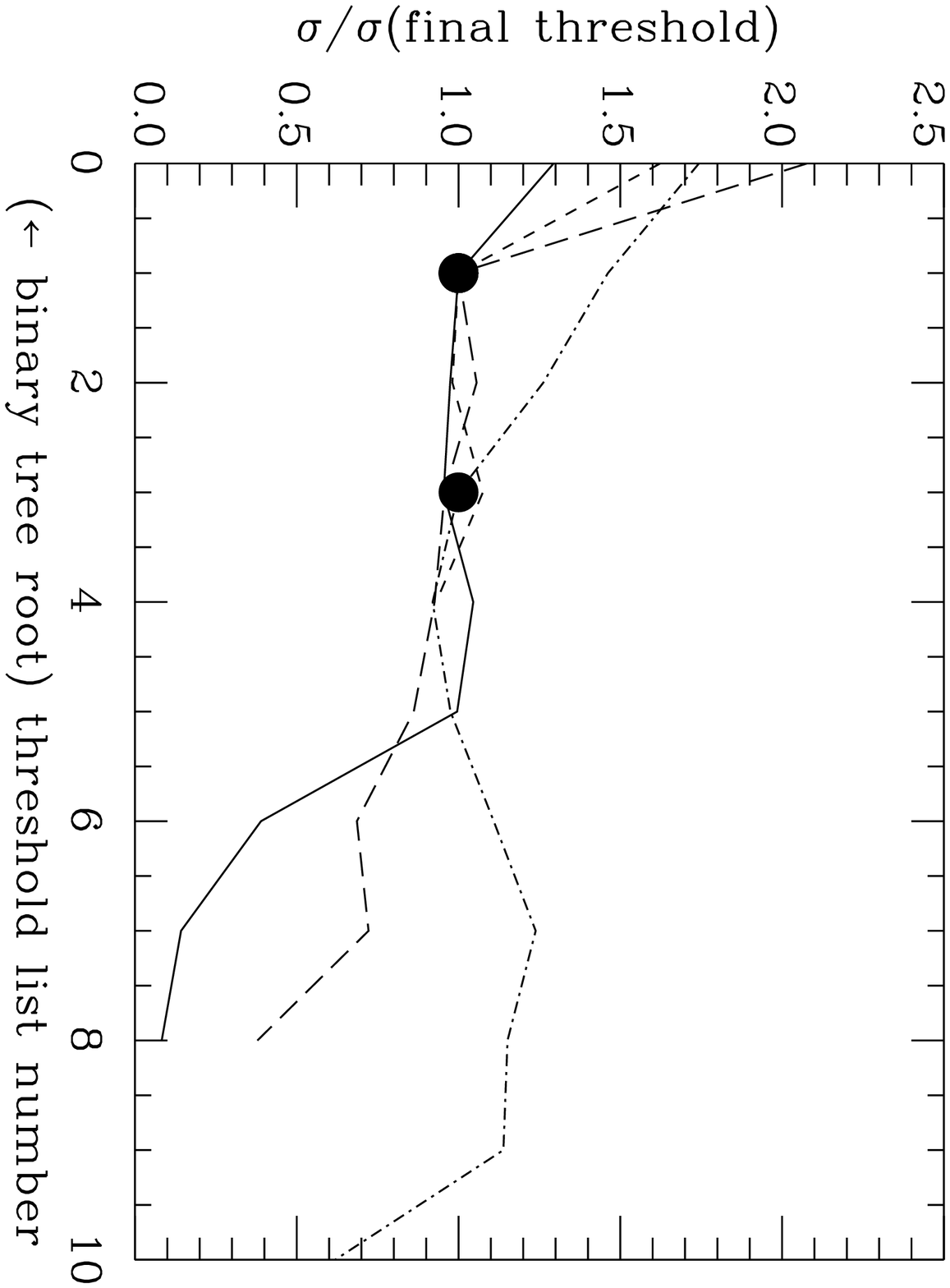}
           {0.25\vsize}              
           {90}                
           {30}                 
           {30}                 
           {100}               
           {-10}                
\caption{Velocity dispersion $\sigma$ computed with the leaves of each threshold
versus the threshold number from the candidate list. The root of the binary tree is on the left
of the abscissa axis. The lines refer to four different lines of sight
of the $\Lambda$CDM cluster shown in Fig. \ref{fig:lambda_rd}. The
velocity dispersion $\sigma$ decreases substantially as we move away from the root,
reaches a plateau and decreases once again.
The starting point of the plateau (filled circles) identifies the
final threshold whose leaves are the cluster members.}
\label{fig:plateau}
\end{figure}

In fact, in our simulated galaxy clusters, $\sigma$ computed with the leaves
of each threshold from our candidate list decreases along the main branch (from the root to the
leaves), reaches a plateau and decreases once again when the main branch splits
into the very internal substructure of the cluster.
The most external threshold of the plateau defines our final threshold.
Fig. \ref{fig:plateau} shows some examples of this trend of $\sigma$.

The leaves of the final threshold are the members of the main cluster. 
Once we have identified the cluster, we can go down in the hierarchy of the main cluster and
find substructure easily. The level of the hierarchy where substructure is
still present gives the degree of their relative binding  energy. 

We can now use the $N$ members of the main cluster to locate the cluster centre.
We compute the two dimensional density distribution
projected onto the sky of the cluster members with the adaptive kernel method
described in Section \ref{sec:fa}, equations (\ref{MM1})--(\ref{MM3}).
The peak of the density distribution determines the celestial coordinates
of the cluster centre. The median of the redshifts of the cluster members
determines the velocity $cz_C$ of the cluster. We prefer the median to the mean
because the median is a more robust estimate of the central value of the velocity distribution.

\section{GALAXY COORDINATES IN THE REDSHIFT DIAGRAM}

The angular separation of each galaxy from the cluster centre
is now trivially (see Fig. \ref{fig:scheme})
\begin{equation}
r_\perp={cz_C\over H_0}\sin\psi.
\end{equation}

The line-of-sight velocity $v_{\rm los}$ requires a more careful consideration.
In real systems, we cannot always separate the peculiar velocity from the Hubble velocity reliably.
The velocity of the galaxy $B$ with respect to the cluster
centre $C$ is ${\bf v}_{B,C}^{\rm pec}={\bf v}_B^{\rm pec}-{\bf v}_C^{\rm pec}=
{\bf v}_B-{\bf v}_C-H_0{\bf r}$, where
${\bf v}_B$ and ${\bf v}_C$ are the physical velocities with respect to the observer
$O$ and ${\bf r}$ is the separation vector $CB$ (Fig.~\ref{fig:scheme}).  We can estimate only
the redshift of the galaxy $B$, $cz_B={\bf v}_B\cdot\hat{\bf r}_{OB}$
and the cluster redshift, $cz_C={\bf v}_C\cdot\hat{\bf r}_{OC}$, where the hat indicates
a versor.  Thus, ${\bf v}_{B,C}^{\rm pec}\cdot\hat{\bf r}_{OB}
=cz_B-cz_C\cos\psi-H_0 r\sin\gamma$
where we have used the fact that ${\bf v}_C\simeq H_0{\bf r}_{OC}$, because
$\vert {\bf v}_C^{\rm pec}\vert\ll H_0\vert{\bf r}_{OC}\vert$.
Finally, the vector ${\bf r}$ is unknown; thus, we define the observable line-of-sight
velocity of galaxy $B$
\begin{equation}
v_{\rm los}\equiv{\bf v}_{B,C}^{\rm pec}\cdot\hat{\bf r}_{OB}
+H_0 r\sin\gamma=cz_B-cz_C\cos\psi.
\end{equation}

This relation yields the desired quantity ${\bf v}_{B,C}^{\rm pec}\cdot\hat{\bf r}_{OB}$
as long as the contribution $H_0 r\sin\gamma$ of the Hubble velocity with
respect to the cluster centre is negligible.
This is not the case when, at fixed angular separation $\psi$, we move $B$ away from
 $A$ (see Fig. \ref{fig:scheme}); in this case both $r$ and $\vert\gamma\vert$ increase and
$v_{\rm los}$ becomes a poor measure of ${\bf v}_{B,C}^{\rm pec}\cdot\hat{\bf r}_{OB}$.
However, $B$ needs to be several Mpc away from $A$ before the Hubble contribution
becomes comparable to the escape velocity which we expect to be
of several hundreds km s$^{-1}$ in a massive cluster (see equation \ref{eq:escnote}).
Moreover, as we move $B$ away from $A$ both the infall (radial) velocity
and the line of sight component of the tangential velocity
decrease, so $v_{\rm los}$ from these galaxies will presumably
be smaller than the $v_{\rm los}$ determining the caustics; thus
the measure of ${\cal A}(r)$ will not be affected.
This problem, however, becomes increasingly serious as we increase $\psi$, because
the relative Hubble contribution to ${\bf v}_{B,C}^{\rm pec}$ increases
and the escape velocity determining the caustic decreases.

Finally, note that the procedure adopted to locate the cluster 
centre also yields, at the chosen threshold, a number of groups distinct from the main
cluster, besides a number of individual galaxies, i.e. galaxies that do not belong
to any group. In the redshift diagram we include only individual galaxies or 
galaxies which belong to groups  with $\pi<cz_{\rm lim}$, where $\pi$ is computed with the
group centre coordinates or the individual galaxy coordinates and
the cluster centre coordinates (equation \ref{eq:proj});
we set $cz_{\rm lim}=4000$ km s$^{-1}$. The group
centre coordinates are computed 
with the same procedure used for the main cluster.


\begin{thebibliography}{}

\bibitem [Bartelmann <1995>]{bartelmann95}
Bartelmann M., 1995,\newblock A\&A, 303, 643
\bibitem [Biviano et al. <1997>]{biviano97}
Biviano A., Katgert P., Mazure A., Moles M., den Hartog R., Perea J., Focardi P., 
1997, A\& A, 321, 84
\bibitem [Blanton et al. <1999>]{Blanton99}
Blanton M., Cen R., Ostriker J. P., Strauss M. A., Tegmark M., 1999,\newblock ApJ, submitted
(astro-ph/9903165)
\bibitem [Carlberg et al. <1996>]{carlberg96}
Carlberg R. G., Yee H. K. C., Ellingson E., Abraham R., Gravel P., Morris S., Pritchet C.
J., 1996,\newblock ApJ, 462, 32
\bibitem [Cen \& Ostriker <1999>]{Cen99}
Cen R., Ostriker J. P., 1999,\newblock ApJ, submitted (astro-ph/9809370)
\bibitem [Couchman <1991>]{couchman91}
Couchman H. M. P., 1991,\newblock ApJ, 368, L23
\bibitem [Couchman, Thomas, \& Pearce <1995>]{couchman95}
Couchman H. M. P., Thomas P. A., Pearce F. R., 1995,\newblock ApJ, 452, 797
\bibitem [Cress <1999>]{cress99}
Cress C. M., 1999,\newblock private communication
\bibitem [Croft, Dalton \& Efstathiou <1999>]{croft99}
Croft R., Dalton G., Efstathiou G., 1999,\newblock MNRAS, 305, 547
\bibitem [den Hartog \& Katgert <1996>]{denhartog96}
den Hartog R., Katgert P., 1996,\newblock MNRAS, 279, 349
\bibitem [Dekel, Burstein \& White <1997>]{dekel97}
Dekel A., Burstein D., White S. D. M., 1997,\newblock in Critical Dialogues in Cosmology,
Princeton, 250th Anniversary, ed. N. Turok (Singapore: World Scientific), p. 175
\bibitem [Diaferio \& Geller <1997>]{diaferio97}
Diaferio A., Geller M. J., 1997,\newblock ApJ, 481, 633 (DG)
\bibitem [Diaferio et al. <1999>]{Diaferio99}
Diaferio A., Kauffmann G., Colberg J. M., White S. D. M., 1999,\newblock MNRAS, in press
(astro-ph/9812009)
\bibitem [Efstathiou, Bond \& White <1992>]{efstathiou92}
Efstathiou G., Bond J. R., White S. D. M., 1992,\newblock MNRAS, 258, 1P
\bibitem [Frederic <1995>]{frederic95}
Frederic J. J., 1995,\newblock ApJS, 97, 259
\bibitem [Frenk et al. <1996>]{frenk96}
Frenk C. S., Evrard A. E., White S. D. M., Summers F. J., 1996,\newblock ApJ, 472, 460 
\bibitem [Fukunaga <1990>]{Fukunaga90}
Fukunaga K., 1990,\newblock Introduction to Statistical Pattern Recognition, Second Edition
(San Diego: Academic Press)
\bibitem [Garnavich et al. <1998>]{garnavich98}
Garnavich P. M., et al. 1998,\newblock ApJ, 493, L53
\bibitem [Geller et al. <1999a>]{geller99a}
Geller M. J., Diaferio A., Kurtz M. J., 1999a,\newblock ApJ, 517, L23 
\bibitem [Geller et al. <1999b>]{geller98}
Geller M. J., et al. 1999b,\newblock in preparation
\bibitem[Gourgoulhon, Chamaraux \& Fouqu\'e <1992>]{Gourgoulhon92}
Gourgoulhon E., Chamaraux P., Fouqu\'e P., 1992,\newblock A\&A, 255, 69
\bibitem[Gurzadyan \& Mazure <1998>]{Gurzadyan98a}
Gurzadyan V. G., Mazure A., 1998,\newblock MNRAS, 295, 177
ed. A. Mazure, F. Casoli, F. Durret, \& D. Gerbal, (Singapore: Word Scientific), 54
\bibitem [Jenkins et al. <1997>]{jenkins97}
Jenkins A., et al. 1997,\newblock in Dark and Visible Matter in Galaxies and Cosmological
Implications, eds. M. Persic \& P. Salucci, ASP Conference Series Vol. 117, p. 348
\bibitem [Kaiser, Squires, \& Broadhurst <1995>]{kaiser95}
Kaiser N., Squires G., Broadhurst T., 1995,\newblock ApJ, 449, 460
\bibitem [Kauffmann et al. <1999a>]{kauffmann98}
Kauffmann G., Colberg J., Diaferio A., White S. D. M., 1999a,\newblock MNRAS, 303, 188 
\bibitem [Kauffmann et al. <1999b>]{kauffmann99}
Kauffmann G., Colberg J., Diaferio A., White S. D. M., 1999b,\newblock MNRAS, in press 
(astro-ph/9809168)
\bibitem [Lombardi \& Bertin <1998>]{lombardi98}
Lombardi M., Bertin G., 1998,\newblock A\&A, 335, 1L
\bibitem [Marzke, Huchra, \& Geller <1994>]{marzke94}
Marzke R. O., Huchra J. P., Geller M. J., 1994,\newblock ApJ, 428, 43
\bibitem[Materne <1978>]{Materne78}
Materne J., 1978,\newblock A\&A, 63, 401
\bibitem [Merrifield <1998>]{merrifield98}
Merrifield M. R., 1998,\newblock MNRAS, 294, 347 
\bibitem[Miyamoto <1990>]{Miyamoto90}
Miyamoto S., 1990,\newblock Fuzzy Sets in Information Retrieval and Cluster Analysis
(Dordrecht: Kluwer Academic Publishers) 
\bibitem [Navarro \& Steinmetz <1997>]{navarro97a}
Navarro J. F., Steinmetz M., 1997,\newblock ApJ, 478, 13
\bibitem [Navarro, Frenk \& White <1997>]{NFW}
Navarro J. F., Frenk C. S., White S. D. M., 1997,\newblock ApJ, 490, 493 (NFW)
\bibitem [Nolthenius, Klypin \& Primack <1997>]{nolthenius97}
Nolthenius R., Klypin A. A., Primack J. R., 1997,\newblock ApJ, 480, 43
\bibitem [Pearce \& Couchman <1997>]{pearce98}
Pearce F. R., Couchman H. M. P., 1997,\newblock New Astronomy, 2, 411 
\bibitem [Pearce et al. <1999>]{pearce99}
Pearce F. R., et al. 1999, ApJL, submitted (atro-ph/9905160)
\bibitem [Perea, del Olmo \& Moles <1990>]{perea90}
Perea J., del Olmo A., Moles M., 1990, A\& A, 237, 319
\bibitem [Perlmutter et al. <1998>]{perlmutter98}
Perlmutter S., et al. 1998,\newblock Nature, 391, 51
\bibitem [Pisani <1993>]{Pisani93}
Pisani A., 1993,\newblock MNRAS, 265, 706
\bibitem [<1996>]{Pisani96}
Pisani A., 1996,\newblock MNRAS, 278, 697
\bibitem [Quintana, Ram\'\i rez, \& Way <1996>]{quintana96}
Quintana H., Ram\'\i rez A., Way M. J., 1996,\newblock AJ, 112, 36
\bibitem [Ramella, Pisani, \& Geller <1997>]{ramella97}
Ramella M., Pisani A., Geller M. J., 1997,\newblock AJ, 113, 483
\bibitem [Reblinsky \& Bartelmann <1999>]{reblinsky99}
Reblinsky K., Bartelmann M., 1999,\newblock A\&A, 345, 1
\bibitem [Reg\"os \& Geller <1989>]{regos89}
Reg\"os E., Geller M. J., 1989,\newblock AJ, 98, 755
\bibitem [Riess et al. <1998>]{riess98}
Riess A. G., et al. 1998,\newblock AJ, 116, 1009
\bibitem [Schmalzing \& Diaferio <1999>]{schmalzing99}
Schmalzing J., Diaferio A., 1999,\newblock in preparation
\bibitem [Schmoldt et al. <1999>]{schmoldt99}
Schmoldt I. M., et al. 1999, AJ, in press (astro-ph/9906035)
\bibitem [Seitz \& Schneider <1996>]{seitz96}
Seitz S., Schneider P., 1996,\newblock A\&A, 305, 383
\bibitem[Serna \& Gerbal <1996>]{Serna96}
Serna A., Gerbal D., 1996,\newblock A\&A, 309, 65
\bibitem [Silverman <1986>]{silverman86}
Silverman B. W., 1986,\newblock Density Estimation for Statistics and Data Analysis (London: Chapman \& Hall)
\bibitem [Strauss \& Willick <1995>]{strauss95}
Strauss M. A., Willick J. A., 1995,\newblock Phys. Rep., 261, 271 
\bibitem [Squires \& Kaiser <1996>]{squires96}
Squires G., Kaiser N., 1996,\newblock ApJ, 473, 65
\bibitem[Tormen, Diaferio \& Syer <1998>]{Tormen98}
Tormen G., Diaferio A., Syer D., 1998,\newblock MNRAS, 299, 728
\bibitem[Tully <1987>]{Tully87}
Tully R. B., 1987,\newblock ApJ, 321, 280
\bibitem [van Haarlem et al. <1993>]{vanhaarlem93a}
van Haarlem M. P., Cay\'on L., de la Cruz C. G., Martin\'ez-Gonz\'alez E., Rebolo R., 
1993,\newblock MNRAS, 264, 71
\bibitem [Van Haarlem \& van de Weygaert <1993>]{vanhaarlem93b}
van Haarlem M. P., van de Weygaert R., 1993,\newblock ApJ, 418, 544
\bibitem [Vedel \& Hartwick <1998>]{vedel98}
Vedel H., Hartwick F. D. A., 1998,\newblock ApJ, 501, 509
\bibitem [Weinberg, Hernquist \& Katz <1997>]{weinberg97}
Weinberg D. H., Hernquist L., Katz N., 1997,\newblock ApJ, 477, 8
\bibitem [Yahil \& Vidal <1977>]{yahil77}
Yahil A., Vidal N. V., 1977, ApJ, 214, 347

\end{thebibliography}
\end{document}